# A First-Principles Comparative Study of Lithium, Sodium, and Magnesium Storage in Pure and Gallium-Doped Germanium: Competition between Interstitial and Substitutional Sites

Fleur Legrain[1] and Sergei Manzhos[1,a]

[1]Department of Mechanical Engineering, National University of Singapore, Block EA #07-08, 9 Engineering Drive 1, Singapore 117576, Singapore

**Abstract**

Thermodynamics and kinetics of Li, Na, and Mg storage in Ge is studied ab initio. The most stable configurations can consist of tetrahedral, substitutional, or a combination of the two types of sites. In the dilute limit, Li, Na prefer interstitial, while Mg prefers substitutional sites. At higher concentrations of Li, Na, and Mg, there is a combination of interstitial and substitutional sites. This is an important finding, as previous ab initio studies of alloying type electrode materials ignored substitutional sites. Insertion energies computed at dilute concentration ($x = 1/64$) show that Na and Mg insertion are not thermodynamically favored in Ge vs formation of bulk Na and Mg, as opposed to Li insertion which is favored. We investigate the effect of *p*-doping of Ge (with Ga) on the thermodynamics and find that it considerably lowers the defect formation energies associated with the insertion of Li/Na/Mg at tetrahedral sites. On the other hand, the energetics associated with Li/Na/Mg insertion at substitutional sites are not significantly affected. In addition, we compute the migration energy barriers for Li/Na/Mg diffusion between two tetrahedral sites (0.38/0.79/0.66 eV), between two substitutional sites (0.77/0.93/1.83 eV), and between two sites of different type (2.15/1.75/0.85 eV).

---

[a] Author to whom correspondence should be addressed; E-Mail: mpemanzh@nus.edu.sg; Tel.: +65-6516-4605; Fax: +65-6779-1459

# 1. Introduction

Electrochemical batteries are believed to be a key technology to decarbonize the transportation sector (by powering electric vehicles) and the electricity supply (by facilitating the integration of intermittent energy sources such as solar or wind into the electric grid). For such applications, Na- [1, 2] and Mg- [3, 4] ion batteries have emerged as promising alternatives to the already well-developed and widely used Li-ion batteries. In particular, Na's abundance is believed to be essential for grid applications, and Mg-ion batteries have the potential for better energy densities (compared to those of Li), which is critical for applications in electric vehicles. We focus here on the negative electrode of Na- and Mg-ion batteries, since an anode material which would provide simultaneously a high energy density, a good rate capability, and a high cycle life, has not yet emerged neither for Na- nor for Mg-ion batteries. Specifically, we investigate germanium which has been shown to be a promising material for Na-ion batteries [5-8] (as well as Li-ion batteries [9-12]). For Mg, however, the performance of Ge has, to the best of our knowledge, not been reported. A paper from year 1979 suggests that intercalation of Mg in diamond Ge is difficult and that alloying may not happen: only a small amount of Mg impurities was observed when Mg was in contact with Ge [13]. Ge provides volumetric capacities which are almost as high as Si's (Si values are given in parenthesis): 2200 mAh/cm$^3$ (2350 mAh/cm$^3$) for Li (Li$_{22}$Ge$_5$), 860 mAh/cm$^3$ (920 mAh/cm$^3$) for Na (NaGe), 2690 mAh/cm$^3$ (2760 mAh/cm$^3$) for Mg (Mg$_2$Ge); these volumetric capacities are computed using the volume of the final state of charge. In addition, compared to Si, Ge provides higher ionic diffusivities [12] and more stable insertion sites [14] for Li/Na/Mg (i.e. intercalation of Na and Mg can be easier in Ge compared to Si), and its isotropic expansion (reported for Li) can reduce the mechanical stresses generated during intercalation compared to the anisotropic swelling of Si [15, 16].

It is well established that Li doping of Si results in interstitial Li defects. This is apparently also true for Na insertion [14, 17, 18]. The Mg impurities observed in Ge in Ref. [13] were attributed to substitutional defects. This suggests that Ge may possess more favorable insertion energetics for substitutional defects than for interstitial. This is important for potential use of Ge in electrochemical batteries, as it would strongly affect both voltages and rate capability. It is therefore important to understand the thermodynamics and kinetics of insertion of Li, Na, and Mg into Ge at different types of sites. We therefore present here a comparative computational study of the insertion and diffusion of Li, Na, and Mg in Ge at different concentrations at substitutional and interstitial sites.

Furthermore, because we have previously shown that *p*-doping could significantly lower the insertion energies of Li, Na, and Mg in Si, which was important as it could facilitates the intercalation of Na and Mg in Si (as these atom do not easily insert into pure Si) [19], we study here whether *p*-doping has a similar effect on Li/Na/Mg insertion in Ge. This is of practical importance, as heavily Ga-doped Ge is achievable (Ga concentration of ~8 at. %) [20].

## 2. Methods

All calculations are performed using density functional theory (DFT) [21, 22] with the Vienna Ab initio Simulation Package (VASP) [23]. The projector augmented wave (PAW) method [24] is employed to describe the ion-electron interactions. We use the PBE functional [25] to approximate the exchange-correlation effects. We model the interaction of Li/Na/Mg with Ge and Ga-doped Ge using simulation cells containing 64 host atoms (Ge and Ga), of size ~11.5×11.5×11.5 Å. The Brillouin zone is sampled with a 6×6×6 Monkhorst-Pack [26] Γ-centered *k*-point mesh. A cutoff of 300 eV is used for the plane wave basis set. The atomic structures and cell vectors are optimized until all forces are below 0.01 eV/Å.

The energy barriers for Li/Na/Mg diffusion between substitutional sites and between the two different types of sites (tetrahedral and substitutional) are computed using the CI-NEB method [27] with 3 images. For those CI-NEB calculations the *k*-point mesh was reduced to 3×3×3. The force tolerance for each image was set to 0.01 eV/Å; for some images along T-T and T-S paths (see below), this force tolerance of was never reached due to numeric instabilities, but the energies were converged.

The energy barriers for the diffusion of Li/Na/Mg between two tetrahedral sites are found by computing directly the transition site (which is the hexagonal site) as this diffusion pathway is well known [14, 28, 29].

The energetics are analyzed by computing the defect formation energies $E_f$:

$$E_f = \frac{1}{n}\big(E(M_n host) - E(host) - nE(M)\big) \qquad \text{(Eq. 1)}$$

In the equation Eq. 1, $M$ stands for Li/Na/Mg, $host$ for Ge or Ga-doped Ge, $n$ represents the number of $M$ atoms, $E(M_n host)$ is the energy of the Li/Na/Mg-inserted Ge (or Ga-doped Ge) structures, $E(host)$ the energy of the host (ideal Ge or the optimized Ga-doped Ge configurations), and $E(M)$ the energy of one atom of Li/Na/Mg in its bulk form (*bcc* for Li and Na, *hcp* for Mg). Negative values for $E_f$ indicate that the insertion of Li/Na/Mg are favored vs. clustering of Li/Na/Mg at the surface of the electrode; on the opposite, positive values indicate that the insertion is not favored. In the configurations where Li/Na/Mg substitutes a Ge atom, the formula of the defect formation energies is adapted to:

$$E_f = \frac{1}{n}\big(E(M_n(host - xGe)) - E(host) - nE(M) - xE(Ge)\big) \qquad \text{(Eq. 2)}$$

In the equation Eq. 2, $x$ is the number of Ge atoms substituted by Li/Na/Mg and $E(Ge)$ is the energy of one atom of Ge in diamond Ge.

The strain energies associated with the different configurations are expressed as [30]:

$$E_s = \frac{1}{n}\big(E(M_n host - M_n) - E(host)\big) \qquad \text{(Eq. 3)}$$

In the equation Eq. 3, $E(M_n host - M_n)$ is the energy of the Ge structure as distorted by the insertion of Li/Na/Mg: from the optimized Li/Na/Mg-inserted Ge structure, the Li/Na/Mg atoms are removed, if some Li/Na/Mg atoms were substituting Ge atoms, they are replaced by Ge atoms; this distorted structure of Ge is then kept fixed and its energy calculated.

We use the Bader charge analysis [31] to compute the charges transferred from Li/Na/Mg to the Ge framework upon insertion/substitution.

## 3. Results and Discussion

### 3.1. Insertion energetics of Li, Na, and Mg in Ge

#### 3.1.1. At low concentration ($x = 1/64$)

We consider different configurations for the insertion of Li, Na, and Mg in the diamond-type crystal lattice of Ge. In particular, we optimized the four following configurations: Li/Na/Mg insertion at a tetrahedral interstitial site (T), Li/Na/Mg insertion at a hexagonal interstitial site (H), Li/Na/Mg substitution of a Ge atom (S), and Li/Na/Mg substitution of a Ge atom with the substituted Ge atom located at a tetrahedral interstitial site neighbor of the Li/Na/Mg atom (S + Ge$_T$). The structures are depicted in Figure 1, and the defect formation energies are given in Table 1. The results show that, among all configurations considered for insertion, inserted Li and Na are more favorable at the tetrahedral interstitial site (T). For Mg, the substitutional defect is preferred (S), which is different from Mg's behavior in Si - in which Mg insertion is more favorable at T sites [13]. The substitutional mechanism for Mg in Ge is in good agreement with experimental work showing that Mg acts as a *p*-dopant in Ge [13].

In the attempt to rationalize the differences in insertion energetics between Li, Na, and Mg at the different sites, we computed the strain energies and the charge transfers associated with the different insertion sites. The strain energies give an estimation of the distortion of the Ge framework upon Li/Na/Mg insertion. On the other hand, the charge transfers (computed using the Bader charge analysis) indicate the amount of charge donated by the Li/Na/Mg atom to the Ge framework. The ionicity of the Ge-Li/Na/Mg bond is believed to reflect the strength of the Ge-Li/Na/Mg interaction as well as the weakening of the Ge-Ge bonds [17, 32]. The results summarized in Table 2 show that the Ge framework is significantly more distorted when Li/Na/Mg inserts at an interstitial site (T or H) than when it inserts at a substitutional site. Among the interstitial sites (T and H), the tetrahedral site is found to lead to a smaller distortion compared to the hexagonal site, rationalizing that all Li, Na, and Mg are more stable at the tetrahedral site than at the hexagonal site. The charge transfers also follow the same trend: the Ge-Li/Na/Mg interaction is greater at the tetrahedral site than at the hexagonal site, even though the differences in charges between T and H are very small. The charge donated by the Li/Na/Mg atom to the Ge framework appears to be notably smaller in the case of substitutional defects than in the case of interstitial defects. We believe that this is due to a difference in insertion mechanism: the valence electrons of Li/Na/Mg, upon their insertion in an interstitial site, are transferred to the conduction band (the density of states plots are given in Figures S1 and S2 in SI), i.e. to the anti-bonding orbitals of Ge, in agreement with previous studies [17, 19, 30]; in contrast, the valence electrons of Li/Na/Mg at a substitutional site are found to be accommodated in the valence band, in the newly $sp^3$ states formed between Ge and Li/Na/Mg (see Figure S1 and S2 in SI). It could therefore be understood that less charge is donated to the Ge framework in the case of a substitutional insertion than that of an interstitial insertion. Regarding the differences between atom types (Li/Na/Mg), we believe that the preference of Li/Na for an interstitial site and that of Mg for a substitutional site can be rationalized by the number of their valence electrons: the insertion of Li/Na in an interstitial site results in a single electron in the anti-bonding orbitals (vs. 2 electrons for Mg); on the other hand, the insertion of Mg in a substitutional site forms only 2 holes in the conduction band (vs. 3 holes for Li/Na).

The defect formation energies given in Table 1 are computed with respect to Li/Na/Mg bulk (*bcc* Li/Na and *hcp* Mg). The results show that among the metal atoms considered (Li/Na/Mg), only Li insertion in Ge is thermodynamically favored (vs. Li/Na/Mg clustering at the surface of the Ge electrode). The defect formation energy for Na and Mg are 0.82 eV and 1.18 eV, respectively. The positive defect formation energy for Na (0.82 eV) may rationalize why Na does not insert easily in crystalline Ge [6]. The highly positive value for Mg insertion in Ge (1.18 eV) may also rationalize the absence of significant intercalation of Mg in Ge [13].

3.1.2. At higher concentrations (*x* = 2/64, 4/64, 8/64)

We also investigate the energetics of Li/Na/Mg insertion in Ge at higher concentrations: 2, 4, and 8 atoms are inserted per simulation cell (of 64 atoms of Ge). We consider the configurations with well-dispersed sites, i.e. by maximizing the M-M interatomic distances (M = Li/Na/Mg). This is because we want to effectively model the different concentrations (x = 2/64, 4/64, 8/64) and not the possible clustering of Li/Na/Mg atoms in Ge. We examine different combinations of interstitial (T) and substitutional (S) sites for Li/Na/Mg insertion. The defect formation energies associated with the different Li/Na/Mg distributions are plotted in Figure 2. The results show that for Li, while the insertion of 1 and 2 Li atoms are more favored at tetrahedral sites, the insertion of 4 and 8 Li atoms are more favored when 3/4$^{th}$ of the Li atoms (i.e. 3 out of 4 atoms and 6 out of 8 atoms) insert at tetrahedral sites while the rest - 1/4$^{th}$ of the Li atoms - substitutes Ge atoms. For Na, for concentrations of up to 4 Na atoms per simulation cell, it is preferred that all Na atoms insert in tetrahedral sites. For the insertion of 8 Na atoms, the lowest configuration involves the substitution of a single Ge atom, in addition to the insertion of all other (i.e. 7) Na atoms at tetrahedral sites. For Mg, while the insertion of 1 atom is more favored when Mg substitutes a Ge atom, the lowest configurations for the insertion of 2 and 4 Mg atoms are found to be those for which half of the Mg atoms insert at substitutional sites and the other half at tetrahedral sites. For 8 Mg, the repartition of lowest energy is composed of 5 substitutional sites and 3 tetrahedral sites. To understand the different repartition of Li/Na/Mg atoms between tetrahedral and substitutional sites, we looked (i) at the electrons / holes which are brought in the conduction / valence band upon Li/Na/Mg insertion, as well as (ii) the defect formation energy of the individual insertion sites (at low concentration) for each Li/Na/Mg. As mentioned earlier, investigation of the density of states shows that upon insertion at tetrahedral site, Li/Na (Mg) brings 1 (2) electron(s) in the conduction band, in agreement with previous studies [17, 19, 30]. On the other hand, when Li/Na (Mg) inserts at a substitutional site, 3 (2) holes are created. In the configurations mixing tetrahedral and substitutional sites, the extra electrons brought by the interstitial Li/Na/Mg are found to fill the holes created by the substitutional Li/Na/Mg. For Li and Mg, the most stable repartitions of tetrahedral and substitutional sites identified for the different concentrations (except 8/64 for Mg) coincide exactly with the balance of the $n/p$-doping: the number of extra electrons brought by the interstitial atoms fill exactly the holes created by the substitutional atoms. It is not surprising that the configurations preferred for Li/Mg-inserted Ge are the ones which maximize the number of electrons in bonding states and minimize the one in anti-bonding states. This is, to the best of our knowledge, the first time this effect is being reported. The substitutional defects are usually ignored in DFT works studying Li (as well as Na and Mg) in alloying type electrode materials [14, 28, 29, 33]. We find here that substitutional defects must play a role. Interestingly, the behavior of Na for insertion is different. For Na, a configuration with a higher number of tetrahedral sites is preferred (i.e. with electrons in the conduction band). This may be because insertion of Na in a substitutional site is particularly unstable (2.42 eV). Similarly for Mg, the substitutional sites which are preferred at low concentration outnumber the tetrahedral sites at a concentration of $x = 8/64$. Therefore, in order to rationalize the different repartitions of tetrahedral and substitutional sites at different concentrations, one needs to account for both the number of holes/electrons in the valence/conduction band and the energetics of the two different types of sites at low concentration. But the most important finding here is that the substitutional sites should not be ignored as it is usually done with group IV alloying materials [14, 28, 29, 33].

Looking at the values of the defect formation energies for the different concentrations, the results show that $E_f$ for Li and Na do not change much with an increase in Li/Na concentration. For Mg, we can see that there is a slight drop in defect formation energies from a concentration of $x = 1/64$ (1.18 eV for 1 Mg in S) to $x = 2/64$ (0.96 eV for 1 Mg in S and 1 Mg in T). We investigate whether this change in $E_f$ could result from an overestimation of $E_f$ for the concentration of $x = 1/64$ because it is not possible to insert half of the Mg atom at S and the other half at T. We therefore consider the configuration corresponding to a concentration of $x = 1/64$ for which the supercell is doubled (in one direction) and two atoms of Mg are inserted, one in a substitutional site and the other in a tetrahedral site. The defect formation energy is found to be 1.14 eV. There is indeed a small decrease in $E_f$ for Mg when going from the concentration $x = 1/64$ to $x = 2/64$. Except this, the defect formation energies for the insertion of Li, Na, and Mg are found to not change significantly in the range of concentrations 1/64…8/64.

*3.2. Migration barriers for the diffusion of Li, Na, and Mg in Ge*

3.2.1. Migration pathways: T-T, S-S, T-S

We now investigate Li, Na, and Mg diffusion in Ge. Because Li, Na, and Mg are found to insert as a combination of tetrahedral and substitutional sites, several diffusion paths need to be considered: the path between two tetrahedral sites (T-T), the one between two substitutional sites (S-S), and the one between the two different types of sites (T-S). The different diffusion paths are shown in Figure 3. The first path (T-T) has been well studied for Li in Si and it is known to happen via hexagonal sites [14, 28, 29]. The second diffusion path (S-S) is modeled by swapping a Li/Na/Mg atom inserted in a substitutional site with a neighboring Ge atom. We model the third path (T-S, i.e. when one Li/Na/Mg atom goes from a tetrahedral to a substitutional site) as a kick-out diffusion mechanism: the Li/Na/Mg atom, initially at a tetrahedral site, substitutes a Ge atom by kicking it off its original position. We therefore model the final state of the diffusion path as the configuration S + Ge$_T$ (Li/Na/Mg substitutes a Ge atom with the substituted Ge atom in a neighboring tetrahedral site). From this configuration (S + Ge$_T$) to a simple substitutional defect (S), the kicked out Ge will need to diffuse away from the defect. The migration barrier for the diffusion of self-interstitials in Ge has been reported for different states of charge of the interstitial Ge atom [34]. The Bader analyses performed on our calculations indicate that the interstitial Ge atom is neutral. For this state of charge, the migration barrier are known to be of about 0.5 eV [34].

The migration barriers computed for each system are given in Figure 3. The lowest migration barriers for all atom types (Li, Na, and Mg) correspond to the T-T migration pathway. This indicates that diffusion of Li, Na, and Mg in (ideal) crystalline Ge mainly happen by diffusion of the metal atoms between tetrahedral sites. The diffusion barriers are found to be 0.38 eV for Li, 0.79 eV for Na, and 0.66 eV for Mg. The energy barrier for Mg, which is high in Si (0.94…1.00 eV [19, 28, 35]) and in particular significantly higher than that of Li in Si (0.56…0.62 eV [17, 19, 29, 36]) is in Ge (0.66 eV) only slightly higher than the value of 0.56…0.62 eV reported for Li in Si [17, 19, 29, 36].

Lithiation of silicon is known to proceed rapidly, and such a barrier therefore means that magnesiation of Ge should not be kinetically hindered. For Na however, although the energy barrier in Ge (0.79 eV) is lower than that in Si (1.06…1.14 eV [14, 17, 35]), it remains relatively high. Diffusion of Li/Na/Mg atoms in the material is important, but it is also necessary that the atoms can diffuse between the two different types of sites (and especially from a T site to a S site) in order to constitute the most stable configurations reported earlier, composed of both tetrahedral and substitutional sites. This is in particular important for Mg, for which the substitutional site allows to lower significantly the insertion energies. For the concentration at which the energy barrier for T-S has been calculated ($x = 1/64$), the energy barrier corresponds (within 0.1 eV) to the energy of the final state, which is that with Li/Na/Mg at an S site and Ge at a neighboring T site (see Figures 1 and 3). For Mg, the migration barrier is 0.85 eV. For Li and Na, the computed migration barriers between T and S sites are particularly high: 2.15 eV and 1.75 eV, respectively. We investigate whether the high migration barriers for Li and Na are due to the fact that at the concentration $x = 1/64$, the configurations S + Ge$_T$ are very unstable for Li and Na (see Table 1). It is indeed only from a concentration of $x = 4/64$ for Li and $x = 8/64$ for Na that the presence of 1 Li/Na is preferred at an S site. We therefore investigated those concentrations. However, for these concentrations the configuration S + Ge$_T$ is found to drift to T upon optimization. We therefore optimize the configuration with S + Ge$_T$ when Ge$_T$ is not at the nearest neighboring site of M$_{Ge}$ but a bit farther away, at a distance of 5.0 Å. The energy difference between that configuration and that in which all Li/Na are in T sites is 2.10 eV for Li and 2.42 eV for Na, suggesting that the T-S barriers are high for Li and Na, even for the concentrations for which the S site starts to be preferred. In addition, as mentioned earlier, in order to form a purely substitutional site S (S + Ge$_T$ is formed with the migration pathway T-S), the Ge atom located at a tetrahedral site Ge$_T$ should diffuse away from the insertion site, which will be done with a migration energy barrier of about 0.5 eV [34].

3.2.2. Effect of dopant-dopant interaction on the T-T migration barriers

For the lowest migration barriers (i.e. T-T), we investigate the effect of Li-Li/Na-Na/Mg-Mg dopant-dopant interaction. More specifically, we compute the energy barriers associated with the diffusion of Li/Na/Mg between two T sites (from $T_i$ to $T_f$) when one Li/Na/Mg atom is located in a nearest neighboring site from $T_i$ (designated by $T_n$), as depicted in Figure 4. We can first compare the energetics of the initial state (dopants in $T_i$ and $T_n$) with that of the final state (dopants in $T_f$ and $T_n$) as well as with that of the configuration with two well-dispersed Li/Na/Mg atoms computed earlier in this work. It was found that the inserted atoms repel each other: the longer the distance between Li/Na/Mg atoms, the more stable the configuration (with an energy difference of 0.04/0.10/0.10 eV between $T_i$-$T_n$ and $T_f$-$T_n$, and of 0.01/0.01/0.02 eV between $T_f$-$T_n$ and the well dispersed sites, for Li/Na/Mg). This can also justify our choice to model well-dispersed configurations to study the concentrations $x =$ 2/64…8/64. Looking at the values of the energy barriers (see Figure 4), we can see that the presence of a neighboring Li/Na/Mg atom reduces considerably the migration barriers between T sites: from 0.38 to 0.28 eV for Li, from 0.66 to 0.52 eV for Mg, and from 0.79 to 0.59 eV for Na.

*3.3. Insertion energetics of Li, Na, and Mg in Ga-doped Ge*

We found previously that the insertion of Na and Mg was thermodynamically not favored in Ge (with highly positive values of $E_f$: 0.82 eV for Na and 1.18 eV for Mg). In order to facilitate (or make possible) the intercalation of Na and Mg in crystalline Ge, we investigate here whether doping Ge with Ga can, similarly to Al in Si [19], lower the insertion energetics for Li, Na, and Mg.

3.3.1. Ga-doped Ge

We study first Ga doping in Ge. We consider the same sites as the ones for Li/Na/Mg: T, H, S, and S + Ge$_T$. The defect formation energies are given in Table 2, and the configurations are identical to the ones depicted in Figure 1. The defect formation energies for Ga insertion in Ge show that Ga atoms prefer to insert at Ge sites (S) and that the insertion of Ga is slightly unfavored ($E_f$ = 0.07 eV). We also consider higher concentrations of Ga doping in Ge that we modeled by maximizing the Ga-Ga inter-distances. The defect formation energies associated with those higher concentrations are also given in Table 3 and show that the defect formation energies slightly increase (Ga insertion is less stable) as the Ga concentration is increased. However, the defect formation energies for a Ga concentration of 8/64 remains close to zero (0.28 eV). This is in agreement with the fact that heavy Ga doping (8 at. %) is achievable in Ge (4/64 and 8/64 correspond to about 6 and 12 at.%) [20].

3.3.2. Li, Na, and Mg insertion in Ga-doped Ge

We investigate the effect of Ga-doping on the energetics for Li, Na, and Mg insertion in Ge. For the concentration in Ga of 1/64, all the concentrations in Li/Na/Mg ($x$ = 1/64…8/64) are considered. We model the configurations for which all Li/Na/Mg are inserted at T sites as well as the most stable combinations of T and S sites identified in pure Ge (or within 0.01 eV in the case of Mg at $x$ = 8/64). For higher concentrations in Ga ($x$ = 2/64…8/64), we only consider dilute concentration in Li/Na/Mg ($x$ = 1/64). The results are presented in Figure 5. They show that the effect of Ga doping on the energetics depends on the type of sites (T or S) of the configurations. When Li/Na/Mg are inserted at T sites, Ga doping considerably lowers the defect formation energies. On the other hand, for the configurations in which Li/Na/Mg are inserted at S sites, Ga doping does not affect much the $E_f$ which remain roughly constant (see rhombuses on the plot for Mg). For the configurations which consist of both T and S sites, the $E_f$ tend to decrease with Ga doping but to a smaller extent compared to the configurations possessing only T sites. As a result, with Ga doping, the configurations which are partially or completely made of S sites tend to become less stable than those entirely made of T sites (or at least made of more T sites). In particular, for Mg at low concentration ($x = 1/64$), the T site becomes more stable than the S site from a concentration in Ga of 1/64. The densities of states corresponding to Ga-doped Ge (given in Figure S3 in SI) show that Ga doping creates electron holes in the valence bond (1 electron hole per Ga inserted). The electron holes are then filled by Li/Na/Mg's valence electrons upon Li/Na/Mg insertion at T sites (while insertion of Li/Na/Mg at S sites results in extra electron holes). The mechanism is similar to that reported for Li/Na/Mg in Al-doped Si [19]. What is particular to Ga-doped Ge is that Ga doping influences the preferred type of defect upon Li/Na/Mg insertion: as Ge is doped with Ga, configurations with T sites are preferred over those with S sites. This may also have the advantage of facilitating the kinetics for Li/Na/Mg diffusion in Ge as the lowest energy barriers correspond to T-T diffusion.

## 4. Conclusions

This study has shown that the lowest energy configurations for Li/Na/Mg insertion consist of both tetrahedral and substitutional sites, with a fraction of substitutional sites of about 1/8 for Na, 1/4 for Li, and 1/2 for Mg. This means that substitutional defects will significantly affect lithiation, sodiation, and magnesiation of Ge and should not be ignored, in particular for Mg. The defect formation energies for the concentrations $x = 1/64...8/64$ are in the range -0.04...0.07 eV for Li, 0.82...1.00 eV for Na, and 0.89...1.14 eV for Mg. This rationalizes why Na and Mg insert less easily in Ge than Li. To facilitate the insertion of Na and Mg in Ge, doping Ge with Ga was shown to be a promising strategy: substituting one atom of Ge by on atom of Ga (concentration of 1/64 or 1.6 at %) lowers the defect formation energies to -0.55...0.00 eV for Li, 0.33...0.79 eV for Na, and 0.74...0.84 eV for Mg. Doping more heavily Ge with Ga lowers even more the insertion energies for Li/Na/Mg. This is a feasible strategy, as doping Ge with up to 8 at.% of Ga was previously achieved. In addition, Ga doping favors T sites over S sites, which may also help the kinetics as the lowest migration barriers are computed for T-T diffusion (0.38 eV for Li, 0.79 eV for Na, and 0.66 eV for Mg), significantly lower than the S-S (0.77 eV for Li, 0.93 eV for Na, and 1.83 eV for Mg) and T-S diffusion (2.15 eV for Li, 1.75 eV for Na, and 0.85 eV for Mg). The energy barriers for T-T diffusion were found to be reduced by Li-Li/Na-Na/Mg-Mg interaction: when one Li/Na/Mg atom is located in a neighboring site of the diffusing Li/Na/Mg atom, the migration barrier is lowered by 0.10/0.20/0.14 eV.

## 5. Acknowledgments

This work was supported by the Ministry of Education of Singapore via an AcRF grant (R-265-000-494-112).

## 7. Tables

Table 1. Defect formation energies (in eV) for Li, Na, and Mg insertion in Ge for different configurations: Li/Na/Mg insertion at a tetrahedral interstitial site (T), Li/Na/Mg insertion at a hexagonal interstitial site (H), Li/Na/Mg substitution of a Ge atom (S), and Li/Na/Mg substitution of a Ge atom with the substituted Ge atom located at a tetrahedral interstitial site neighbor of the Li/Na/Mg inserted atom (S + Ge$_T$).

| $E_f$ | T | H | S | S + Ge$_T$ |
|---|---|---|---|---|
| Li | -0.04 | 0.35 | 1.48 | 2.11 |
| Na | 0.82 | 1.62 | 2.42 | 2.48 |
| Mg | 1.52 | 2.18 | 1.18 | 2.29 |

Table 2. Strain energies (in eV) and charges transferred from Li/Na/Mg to the Ge framework (in $e$) associated with the different configurations considered for Li/Na/Mg-doped Ge at a Li/Na/Mg concentration of $x = 1/64$ (T, H, and S).

| | T | H | S |
|---|---|---|---|
| $E_{strain}$ (eV) | | | |
| Li | 0.15 | 0.29 | 0.07 |
| Na | 0.46 | 0.91 | 0.60 |
| Mg | 0.48 | 0.96 | 0.13 |
| Charges donated by Li/Na/Mg to the Ge framework ($e$) | | | |
| Li | 0.83 | 0.82 | 0.77 |
| Na | 0.72 | 0.71 | 0.66 |
| Mg | 1.51 | 1.50 | 1.06 |

Table 3. Defect formation energies (in eV) for the insertion of Ga in Ge at a Ge site (S) as well as at a tetrahedral (T) and a hexagonal (H) interstitial sites. For concentrations of 2/64, 4/64, and 8/64, well-dispersed Ga distributions are chosen.

| | T | H | S | S + Ge$_T$ |
|---|---|---|---|---|
| 1/64 | 1.98 | 2.95 | 0.07 | 2.20 |
| 2/64 | | | 0.11 | |
| 4/64 | | | 0.20 | |
| 8/64 | | | 0.28 | |

## 8. Figure captions

Figure 1. Configurations associated with Li/Na/Mg inserted at a tetrahedral interstitial site (T), with Li/Na/Mg inserted at a hexagonal interstitial site (H), with Li/Na/Mg substituting a Ge atom (S), Li/Na/Mg substituting a Ge atom with the substituted Ge atom in a neighboring tetrahedral site (S + Ge$_T$). Ge atoms are in grey, Li/Na/Mg atoms in green, and the Ge atom localized at the tetrahedral site in dark grey. Only part of the simulation cell is shown.

Figure 2. Defect formation energies (in eV) for Li/Na/Mg-doped Ge systems at different Li/Na/Mg concentrations ($x$ = 1/64 – empty circles, 2/64 – circles with one vertical bar, 4/64 – circles with two perpendicular bars, 8/64 – full circles) and for different repartitions of tetrahedral and substitutional sites. The defect formation energies are plotted against the fraction of substitutional sites. The lowest configuration for each concentration is indicated with a black dot.

Figure 3. Migration energy curves for Li/Na/Mg diffusion along the different pathways considered: T-T (top left), S-S (top right), and T-S (bottom left). On the bottom right is shown the 64-atom supercell with the different pathways for Li/Na/Mg diffusion. The Li/Na/Mg atoms are in green, the displaced Ge atoms in dark grey, and the other Ge atoms in grey.

Figure 4. Migration energy curves for Li/Na/Mg diffusion between two T sites (from $T_i$ to $T_f$) with (in solid line) and without (in dashed line) the presence of a spectator Li/Na/Mg atom ($T_n$) located in a neighboring T site of $T_i$. The supercell with the migration path and the spectator atom is also shown.

Figure 5. Defect formation energies (in eV) associated with Li/Na/Mg insertion in pure and Ga-doped Ge. Different concentrations of Ga are considered (on the $x$ axis) as well as different concentrations of Li/Na/Mg ($x$ = 1/64 – empty circles, 2/64 – circles with one vertical bar, 4/64 – circles with two perpendicular bars, 8/64 – full circles). The configurations include different repartitions of tetrahedral and substitutional sites. The symbols indicate the fraction of substitutional sites $f_s$ ($f_s$ = 0 – circles, $f_s$ = 1/8 – triangles rotated by 90º, $f_s$ = 1/4 – triangles rotated by 180º, $f_s$ = 1/2 – triangles, $f_s$ = 1 – rhombuses).

## 9. Figures

Figure 1.

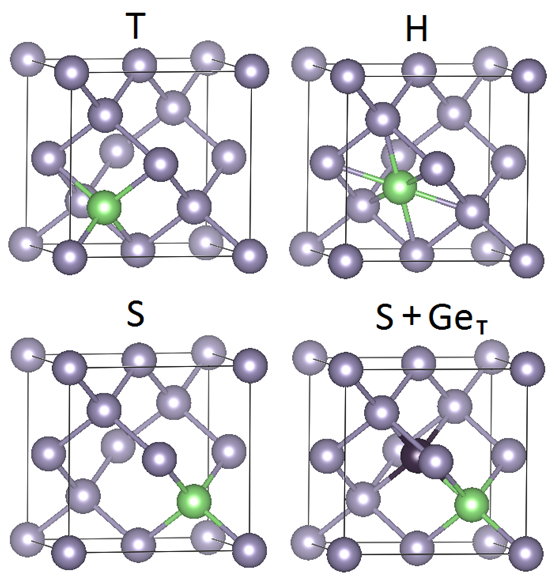

Figure 2.

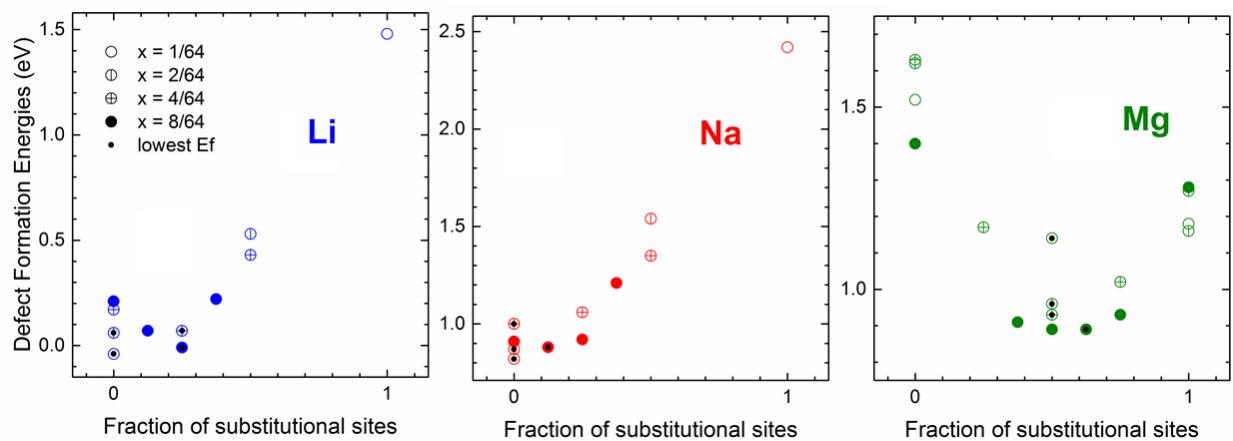

Figure 3.

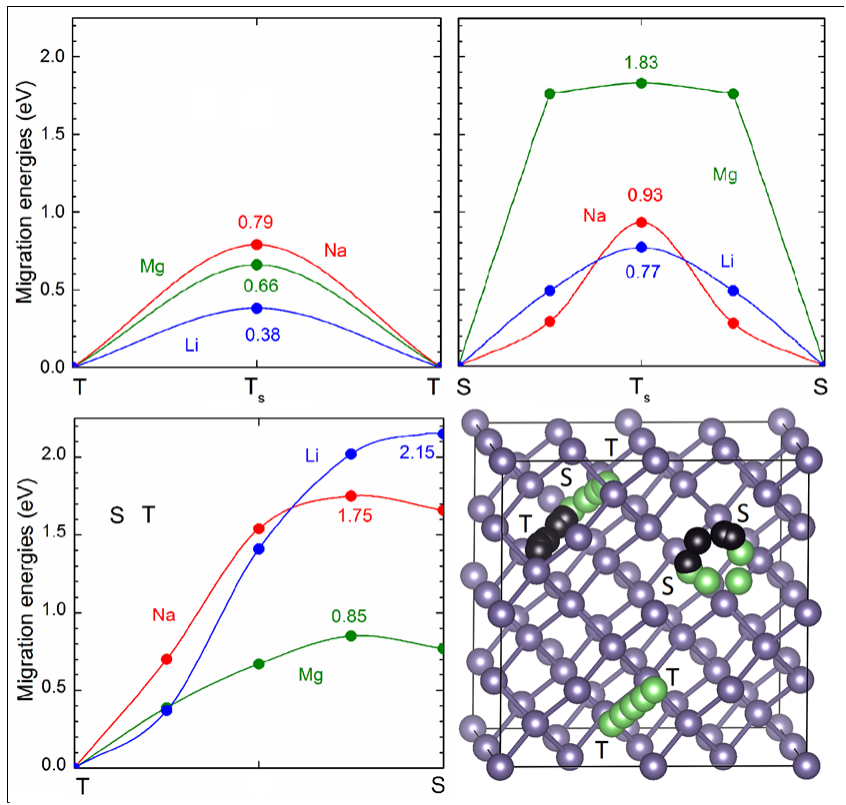

Figure 4.

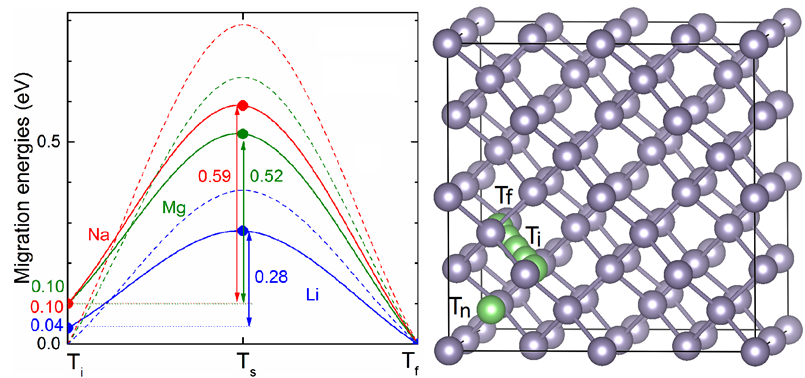

Figure 5.

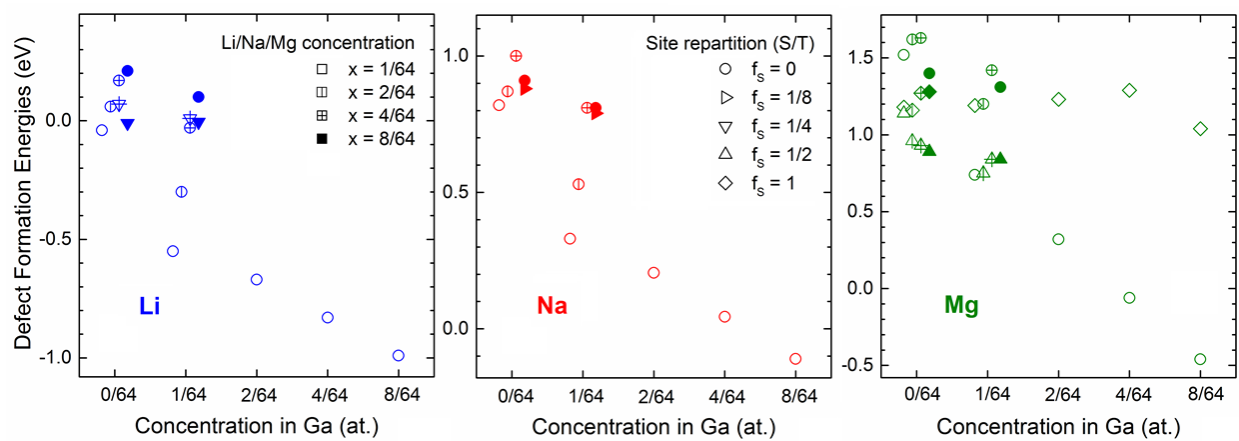